# Spin-injection and spin-relaxation in *p*-doped InGaAs/GaAs quantum-dot spin light emitting diode at zero magnetic field

Alaa E. Giba[1,2], Xue Gao[1], Mathieu Stoffel[1*], Xavier Devaux[1], Bo Xu[3], Xavier Marie[4], Pierre Renucci[4], Henri Jaffrès[5], Jean-Marie George[5], Guangwei Cong[6], Zhanguo Wang[3], Hervé Rinnert[1] and Yuan Lu[1*]

[1]*Institut Jean Lamour, UMR 7198, CNRS-Université de Lorraine, Campus ARTEM, 2 Allée André Guinier, BP 50840, 54011 Nancy, France*
[2]*National Institute of Laser Enhanced Sciences, Cairo University, Giza 12613, Egypt*
[3]*Key Laboratory of Semiconductor Materials Science, Institute of Semiconductors, Chinese Academy of Sciences, P.O. Box 912, Beijing 100083, China*
[4]*Université de Toulouse, INSA-CNRS-UPS, LPCNO,135 Avenue de Rangueil, F-31077 Toulouse, France*
[5]*Unité Mixte de Physique, CNRS, Thales, Université Paris-Saclay, 91767, Palaiseau, France*
[6]*Research Institute for Advanced Electronics and Photonics, National Institute of Advanced Industrial Science and Technology, Onogawa 1-16, Tsukuba, Ibaraki 305-8569, Japan*
*Corresponding authors: *yuan.lu@univ-lorraine.fr; mathieu.stoffel@univ-lorraine.fr*

**ABSTRACT**

We report on efficient spin injection in *p*-doped InGaAs/GaAs quantum-dot (QD) spin light emitting diode (spin-LED) under zero applied magnetic field. A high degree of electroluminescence circular polarization ($P_c$) ~19% is measured in remanence up to 100K. This result is obtained thanks to the combination of a perpendicularly magnetized CoFeB/MgO spin injector allowing efficient spin injection and an appropriate *p*-doped InGaAs/GaAs QD layer in the active region. By analyzing the bias and temperature dependence of the electroluminescence circular polarization, we have evidenced a two-step spin relaxation process. The first step occurs when electrons tunnel through the MgO barrier and travel across the GaAs depletion layer. The spin relaxation is dominated by the Dyakonov-Perel mechanism related to the kinetic energy of electrons, which is characterized by a bias dependent $P_c$. The second step occurs when electrons are captured into QDs prior to their radiative recombination with holes. The temperature dependence of $P_c$ reflects the temperature induced modification of the QDs doping, together with the variation of the ratio between the charge carrier lifetime and the spin relaxation time inside the QDs. The understanding of these spin relaxation mechanisms is essential to improve the performance of spin LED for future spin optoelectronic applications at room temperature under zero applied magnetic field.

# I. INTRODUCTION

Spin light emitting diodes (spin-LED) [1,2] and spin-lasers [3] have attracted intensive interests since two decades because these devices can convert carrier spin polarization to circular polarization of light. Numerous applications are now being explored based on this technology, such as optical telecommunication with circular polarization [4,5], three dimensional (3D) display screens [6], compact circular light sources for biologic applications [7] and quantum cryptography with circular polarized light emitted from a single quantum dot [8] *etc*. The spin-LED devices consist of two parts: the ferromagnetic layer (spin-injector) which is used to inject spin-polarized electrons (or holes) and the semiconductor part in which charge carrier recombination takes place leading to the emission of circularly polarized light. To detect the optical circular polarization with a Faraday geometry (surface detection), the common way is to use a quantum well (QW) based LED structure according to the optical selection rule [9,10]. This allows in principle to obtain 100% of luminescence circular polarization if we inject 100% spin-polarized electrons in the LED structure (if spin relaxation is neglected in the semiconductor part). Low dimensional structures such as quantum dots (QDs) also appear interesting because the spin relaxation time was reported to be much larger than the one in QWs [11,12]. In particular, the doping of QDs plays an important role as it allows the control of the electron-hole exchange interaction [13]. When an electron is injected into a positively charged QD, the formation of a trion ($X^+$) minimizes the exchange interaction between electron and hole, which can result in long spin relaxation time of electrons at low temperature (limited by the interaction of electrons with fluctuating nuclear spins) [12]. Several works have reported the electric spin-injection in either InAs or InGaAs QD based spin-LED [14,15,16,17,18] or spin-laser [19]. A significant electroluminescence (EL) circular polarization (~5%) was measured even at room temperature with a magnetic field of 3T [14].

The CoFeB/MgO spin injector has shown very high spin injection efficiencies [20,21]. By using an ultrathin CoFeB/MgO structure, we have been able to obtain a perpendicularly magnetized spin injector, leading to the observation of a circular polarization at zero applied magnetic field [22,23,24], which is important for practical application. By using such a spin injector, a circular polarization as high as 35% for spin injection in single QD was demonstrated [25]. The high electron spin injection also results in an efficient nuclear spin polarization. This has been identified from the nuclear spin field induced splitting in EL between the two circularly polarized states at zero magnetic field [25].

In this work, by taking benefits from such large circular polarization at zero magnetic field, we investigate the detailed spin injection and spin relaxation mechanisms in such state-of-art sample. There are two key advantages for this kind of structure. Firstly, measurements without magnetic field prevents any magnetic field effect related to the Zeeman splitting [26] and the spin relaxation mechanisms in the semiconductor part [18]. Secondly, long spin relaxation time in QDs can allow us to decouple the spin relaxation process inside and outside of QDs. Here, the bias and temperature dependence of the circular polarization have been systematically studied. We have clearly evidenced a two-step spin relaxation process. The first step is related to the spin relaxation before electrons are trapped into QDs and the second step occurs inside the QDs before the recombination with holes. The understanding of spin-relaxation mechanisms will help to optimize the spin-LED structure to achieve a better performance at room temperature and zero magnetic field.

## II. EXPERIMENTAL DETAILS

The spin-LED structure contains two parts: one part is the semiconductor LED part and the other is the metallic spin injector part. **Fig. 1** schematically shows the structure of the QD based spin-LED. The *p-i-n* LED device grown by MBE contains a single layer of $In_{0.3}Ga_{0.7}As$ quantum dots embedded in the active region. The full sequence of the structure is as following: *p*+-GaAs:Zn (001) substrate ($p=3\times10^{18}$ cm$^{-3}$)/ 300 nm *p*-GaAs:Be ($p=5\times10^{18}$ cm$^{-3}$)/ 400 nm *p*-$Al_{0.3}Ga_{0.7}As$:Be ($p=5\times10^{17}$-$5\times10^{18}$ cm$^{-3}$)/ 30 nm *i*-GaAs with Be δ-doping in the center/ 7ML InGaAs QD/ 30 nm *i*-GaAs/ 50 nm *n*-GaAs:Si ($n=1\times10^{16}$ cm$^{-3}$). The bottom inset of **Fig. 1** shows an AFM image of the InGaAs/GaAs QD layer before capping with GaAs layer. The InGaAs QDs have a density of $1.6\times10^{10}$ cm$^{-2}$. The average lateral dot diameter is about 30 nm and the height is about 9 nm. The Be δ-doping concentration has been calibrated to yield approximatively one hole per dot. The LED was passivated with arsenic in the MBE chamber. Subsequently, the structure was transferred through air into a second MBE-sputtering interconnected system. The As capping layer was desorbed at 300°C in the MBE chamber and then the sample was transferred through ultra-high vacuum to a sputtering chamber to grow the spin injector. The structure of the spin injector is as following: 2.5 nm MgO/ 1.1 nm $Co_{0.4}Fe_{0.4}B_{0.2}$/5 nm Ta. To fabricate the devices, 300μm diameter circular mesas were processed using standard UV photolithography and etching techniques. Finally, the processed wafers were cut into small pieces to perform rapid temperature annealing (RTA) at 300°C for 3 min to establish the perpendicular magnetic anisotropy (PMA) of the spin injector. More details concerning the growth and the optimization of the perpendicular spin-injector can be found in Refs [22,24].

## III. RESULTS AND DISCUSSIONS

*1. Structural characterization of the spin injector*

High resolution scanning transmission electron microscopy (HR-STEM) was performed to characterize the interfacial structure of the spin injector after RTA treatment at 300°C. The images were taken using a JEOL ARM200 cold field-emission gun operating at 200 KV. As shown in the top inset of **Fig. 1**, both MgO/GaAs and MgO/CoFeB interfaces appear sharp. A thin amorphous GaAs layer (~0.4 nm) is detected at the MgO/GaAs interface, which could be due to the large kinetic energy of the atoms reaching the surface during the sputtering growth process. The Ta layer remains almost amorphous after annealing. The MgO layer displays a good (001) textured structure and 11 atomic planes can be well distinguished. The high quality MgO layer serves as a template to crystalize the above CoFeB layer during annealing. The ultrathin CoFeB layer (1.1 nm) is found to be homogenous and partially crystallized to bcc structure with an epitaxial relationship MgO[100](100)//CoFe[110](010). This result is in a good agreement with previous reports concerning the CoFeB crystallization on magnetic tunnel junctions (MTJs) [27,28]. Since the perpendicular magnetic anisotropy properties come from the interface anisotropy created by the hybridization of Fe(Co) and O atoms [22], the precise control of MgO/CoFeB interface structure and chemistry is essential to obtain PMA and high spin injection efficiency [24].

## 2. Magnetic field dependence of circular polarization

For the electroluminescence measurements, the spin-LED was mounted in a cryostat (developed by Cryoscan), which can be cooled down to 10K with liquid helium. The EL signal was detected in a Faraday geometry with the magnetic field applied perpendicular to the sample plane. The detected EL signal was analyzed by a spectrometer (*i*HR 320) with 300 grooves per mm and detected by a Si-based CCD camera. The EL circular polarization $P_c$, which is defined as $P_c=(I^{\sigma+}-I^{\sigma-})/(I^{\sigma+}+I^{\sigma-})$ where $I^{\sigma+}$ and $I^{\sigma-}$ are the intensities of the right and left circularly polarized components of the luminescence, respectively, was analyzed through a $\lambda/4$ wave plate and a linear polarizer.

The insets of **Fig. 2** show the typical polarization resolved EL spectra measured from a spin-LED at 10K under a bias ($V_{bias}$) of 2.5V with a very small current of 6µA (current density of $1.4\times10^{-3}$A/cm$^2$) at $B$=0T. In these spectra, we can observe a single peak centered at about 1.368 eV (wavelength of 906 nm) corresponding to the X$^+$ trion (one electron and two holes forming a singlet) transition for the quantum dot ensemble. Due to the limitation of our spectrometer, the resolution in our spectra is not high enough to distinguish the EL signal originating from single quantum dot as published in Ref. [25]. The left and right insets correspond to the measurements where the sample magnetization was firstly saturated by positive and negative 0.35T field, respectively. In both cases, we measured a large difference of the EL intensities for the right ($I^{\sigma+}$) and left ($I^{\sigma-}$) circularly polarized components at zero field. The EL $P_c$ can be determined to be about +18% and -19% in the left and right insets, respectively. The sign change of $P_c$ is due to the reversal of the magnetization direction of the spin injector. Note that for such a thin CoFeB layer, the spurious component due to the magnetic circular dichroism (MCD) that could artificially increase the measured $P_c$ is estimated to be less than 1%, based on measurements performed in Ref. [22]. We have further examined the evolution of the EL circular polarization as a function of the magnetic field. As shown in **Fig. 2**, the behavior of $P_c$ as a function of magnetic field shows clear hysteresis loop feature, which can fairly match the magnetization hysteresis loop measured on the unpatterned sample by superconducting quantum interference device (SQUID). This gives a strong argument that the large remanent $P_c$ observed at $B$=0T is due to the injection of spin-polarized charge carriers from the ultrathin CoFeB layer with PMA. The high $P_c$ measured at zero magnetic field in the QD spin-LED is almost twice larger than our previous reported $P_c$ values in QW based spin-LED with the same type of PMA spin injector [24], and also approaches the high $P_c$ values reported with *in-plane* magnetized spin injectors [29,30,31]. This high quality sample can be considered as a model system to study spin injection and relaxation mechanisms under zero magnetic field.

3. *Bias dependence of the circular polarization*

In the following, we firstly investigate the bias dependent EL circular polarization at 10K. The inset of **Fig. 3(a)** presents a typical *I-V* characteristic of the spin LED device measured at 10K. The current starts to increase when the bias exceeds 2V. **Fig. 3(a)** shows the evolution of the EL intensity as a function of the injected current. For low injected currents (<25 µA), the EL intensity first increases linearly. However, for currents larger than 25 µA, the EL intensity deviates from a linear behavior. The external quantum efficiency (EQE), which is defined by the ratio of the number of photons emitted from the QDs to the number of electrons passing through the device, decreases with the increasing bias. This can be explained by the filling of QD ground states under high current injection (saturation) [32]. As a consequence, more and more electrons contribute to radiative or non-radiative recombinations at larger bias outside of the QD active region. **Fig. 3(b)** presents the EL spectra as a function of applied bias. The most important feature is that the shape of the spectra remains almost unchanged for different applied biases. The position of the EL peak maxima as indicated by the dashed line are always the same. Moreover, the full width at half maximum (FWHM) of the spectra also remains constant at about 54 meV. All these features indicate that the applied bias does not induce a quantum-confined Stark effect (QCSE) for the light emission from QD [33]. Polarization-resolved spectra were acquired at different bias and the circular polarizations are extracted and presented in **Fig. 3(c)** as a function of applied bias and current. It is found that $P_c$ increases firstly with bias from 11% at 2V to a maximum of 19% at 2.5V and then decreases again to 11% at 4V. The variation with current shows a much sharper increase of $P_c$ to the maximum, which is reached for a small current of 6μA.

In order to understand the bias dependence, we have performed a simulation of the band structure of the spin-LED device by using a self-consistent "1D Poisson" program [34,35]. The nominal layer thicknesses and the dopant concentrations are input parameters for the simulation. For simplification, the $In_{0.3}Ga_{0.7}As$ QD layer has been modeled by a 9 nm thick $In_{0.3}Ga_{0.7}As$ QW layer. As shown in **Fig. 3(d)**, for applied biases below 1V, we found that the applied voltage mainly flattens the band in the GaAs depletion layer, leading to a flat band condition in the LED. When the bias increases, the voltage starts to be applied on the MgO barrier without modifying the band structure of the semiconductor part. For all the bias range considered here, the band structure of the $In_{0.3}Ga_{0.7}As$ active layer is almost unchanged, which is consistent with the fact that the EL peak position remains unchanged for different applied biases (**Fig. 3(b)**). The left inset of **Fig. 3(d)** shows a linear increase of the calculated voltage across the MgO barrier when the applied bias is larger than 1V. The voltage across the MgO results in a band bending in the MgO barrier, which has two consequences. Firstly, the Fermi level of the metallic injector is effectively lifted to be higher than the conduction band minimum of GaAs, thus leading to the appearance of electroluminescence with a very small current injection [20,36]. Secondly, for larger applied biases, the injected electrons have a large kinetic energy after tunneling through the MgO barrier, as schematically shown in the right inset of **Fig. 3(d)**. Concerning the EL circular polarization, the initial increase of $P_c$ is mainly due to the reduction of the electron transit time in the GaAs depletion layer. As reported for the electron spin injection in Si, the electron transit time can vary by two orders of magnitude from diffusion to drift regime [37]. The reduction of the transit time can effectively reduce the possibility of spin flipping due to the spin-relaxation mechanisms, mainly Dyakonov-Perel (DP) mechanism [38], before electrons reach the active region. Therefore, the circular polarization increases for small applied biases. However, when the bias is too large, the electrons having high kinetic energy will quickly lose their spin when they are thermalized to the conduction band. The rate of spin flipping $\tau_s^{-1}$ is proportional to $E_K^3$ [39]. Therefore, the circular polarization decreases for large applied biases. The non-monotonic variation of $P_c$ reflects mainly the spin relaxation mechanism before the electrons reach the active region. It is worth to mention that for thin Schottky barrier based spin-LED with a highly *n*-doped semiconductor layer at the interface, it was demonstrated that a very significant part of the bias drops in the semiconductor active zone. In that case, the dependence of circular polarization on bias is usually explained by the complex bias dependence of the ratio $\tau/\tau_s$ between the carrier lifetime $\tau$ and the electron spin relaxation time $\tau_s$ in the active zone [31,40]. According to our simulations and experiments, this explanation does not hold for CoFeB/MgO spin injector as the "flat band" conditions remain even for large bias.

## 4. *Temperature dependence of the circular polarization*

Let us now focus on the temperature dependence of the circular polarization. **Fig. 4(a)** shows the EL intensity as a function of temperature measured for two fixed injected currents of 6μA and 30μA. The EL intensity decreases drastically with increasing temperature for both injection currents. Above 110K, the EL intensity becomes too low to be detected. This drop of EL intensity is due to the thermal escape of carriers out of the QDs [41,42]. Here, this well-known effect is very pronounced since the InGaAs QDs emission peak (~1.368 eV) in **Fig. 3(b)** is very close to the wetting layer emission one (~1.409 eV), *i.e.* the dots are characterized by rather weak confinement energies compared to the self-organized quantum dots usually investigated [43].

The polarization-resolved spectra were then acquired to study the temperature dependence of $P_c$, as presented in **Fig. 4(c)** for both 6μA and 30μA injection currents. In comparison to the monotonic decrease of the EL intensity, the evolution of $P_c$ exhibits non-linear variation with temperature: firstly $P_c$ decreases to reach a minimum at about 80K and then $P_c$ increases from 80 K up to 110K. Below 60K, $P_c$ measured for a 6μA injection current is larger than $P_c$ measured for a 30μA injection current. For temperatures beyond 60 K, the opposite behavior is observed. To understand this temperature dependence, we have measured the *I-V* characteristics of the spin-LED device as a function of temperature (**Fig. 4(b)**). The junction resistance, which consists of the MgO barrier and the semiconductor part, shows a decreasing behavior when the temperature increases from 10 K to 100

K. For a constant injection current, the bias applied on the device decreases when the temperature increases. Therefore, the temperature dependence of $P_c$ at *a fixed current* is coupled with the bias dependence of $P_c$. To better understand the temperature dependence, we have further measured the bias dependence of $P_c$ at three different temperatures (10K, 60K and 100K), as shown in **Fig. 4(d)**. The bias dependences at 10K and 100K show similar behavior. The small shift of the optimum bias to 2.75V at 100K is due to a small change of the resistance ratio between the MgO barrier and the semiconductor part at higher temperature. In contrast, $P_c$ depends less on the applied bias at 60 K. This agrees with the observation that $P_c$ values are almost the same for 6μA and 30μA injection current at 60K (**Fig. 4(c)**). As discussed above, the bias dependent $P_c$ is mainly related to the band structure change in the MgO barrier and GaAs depletion region. Therefore, studying the temperature dependence of $P_c$ at *a fixed voltage* can reveal the intrinsic spin relaxation mechanism in the QD. **Fig. 4(e)** shows the polarization-resolved EL spectra measured at 10K, 60K, and 100K at $V_{bias}$=2.75V. With increasing temperature, the EL peak position shifts to lower photon energy and meanwhile the FWHM of the EL spectra increases. The measured $P_c$ firstly decreases from 18% at 10K to 9% at 60K, then increases again to 19% at 100K. The increase of $P_c$ at higher temperature is much more pronounced than the measurement done at a fixed injection current. In **Fig. 4(d)**, the dashed black and red lines intuitively show the evolution of $P_c$ with temperature for fixed injection currents of 6μA and 30μA, respectively, which allows us to easily understand the temperature dependent behavior shown in **Fig. 4(c)**.

Since the voltage partition between MgO and the semiconductor part is almost the same at 2.75V by varying temperature, the temperature dependence of the circular polarization should unveil the spin dynamics which occur after electrons are captured into QD and before charge carrier recombination. The EL circular polarization ($P_c$) can be related to the electron spin-polarization injected into QDs ($P_{inj}$) as:

$$P_c = \alpha \cdot F \cdot P_{inj} \qquad (1),$$

where $F$ is expressed as $F = \frac{1}{1+\tau/\tau_s}$ [44], $\tau$ and $\tau_s$ are the carrier lifetime and spin relaxation time inside QDs, respectively. The parameter $\alpha$ (varying from 1 to 0) is linked to the average doping per quantum dot, which varies significantly when the temperature increases. This variation of QD doping was clearly identified in the temperature dependence measurements of inter-sublevel absorption [45]. It has strong consequences on the amplitude of the measured EL circular polarization since the QD eigenstates change with the doping. It is well known that the exciton eigenstates of undoped InGaAs quantum dots are linearly polarized as a consequence of the anisotropic exchange interaction [46]. As a consequence, injecting electrically a spin polarized electron inside such a neutral quantum dot will not lead to the emission of circularly polarized luminescence in a continuous wave (CW) measurement, *i.e.* $\alpha \sim 0$ [11,47]. However, if the QD contains a resident hole, the EL emitted by the Spin-LED will reflect the spin polarization of the injected electron spin, *i.e.* $\alpha \sim 1$. Indeed, the eigenstates of the charged exciton (one electron and two holes forming a singlet, as shown in the inset of **Fig. 4(a)**) are circularly polarized: the exchange interaction between electron and holes vanish [46]. This means that the average *p*-doping per QD in the Spin-LED device plays a crucial role on the amplitude of the measured circularly polarized EL independently of the electron and spin lifetimes.

The decrease of the EL circular polarization from 10 to 80 K observed in **Fig. 4(c)** can be explained by Eq. (1). In this temperature range, the investigations in spin-LED based on quantum wells show that $P_{inj}$ varies very weakly [22]. Moreover, it was shown that the electron spin relaxation time $\tau_s$ from a single InAs QD trion is also almost constant from 0 to 50 K [48]. As a consequence, the $P_c$ decrease in **Fig. 4(c)** can be related either to a increase of the lifetime $\tau$ or a decrease of $\alpha$ due to the doping variation of the dot. The lifetime increase in QD is very unlikely when the temperature increases. Thus we interpret the drop of $P_c$ as a consequence of the reduction of $\alpha$ (optimized to be close to 1 at 10 K). This interpretation is fully consistent with the EL quenching presented in **Fig. 4(a)** which results from the thermal escape of the holes from the QDs. For temperatures above 80 K, we observe an increase of $P_c$ in **Fig. 4(c)**. This could be due to a decrease of the electron lifetime resulting from the electron escape from the quantum dots. Since the energy difference for electrons between the QD ground state and the wetting layer states is larger compared to the one for holes (see inset of **Fig. 4(a)**), it explains that this effect occurs at higher temperatures [45,49].

By investigating the spin-injection in a nominally *undoped* InGaAs QD spin LED, C. H. Li *et al.* [14] reported that $P_c$ (~5% for $B$=3T) is almost constant and can be measured even at room temperature. The difference with our work could rely on the characteristics of the InGaAs quantum dots characterized in this reference by much weaker thermal escape of carriers thanks to a better confinement of carriers (emission wavelengths of the quantum dots beyond 1 μm).

5. *Two-step spin relaxation process*

An important result from this work is the clear evidence of a two-step spin relaxation process. The first step occurs when electrons tunnel through the MgO barrier and travel across the GaAs depletion layer, which is manifested by a bias dependent $P_c$ measurement. The second step occurs when electrons are captured into QDs prior to their recombination with holes, which is manifested by a temperature dependent $P_c$ measurement. The two steps can be decoupled and expressed by the following relationship:

$$P_c = \alpha \cdot F \cdot P_{inj} = \gamma(T) \cdot \eta(V) \cdot P_e \qquad (2),$$

where $\eta$ characterizes the spin relaxation in the first step (which depends on the bias) and $\gamma$ characterizes the spin dynamics in the QD (which depends on the temperature). $P_e$ represents the spin injection efficiency when electron enters into the semiconductor part. Let us focus on four different $P_c$ values measured at 10K and 60K and under applied biases of 2V and 2.5V. At 10K and 2V, the spin relaxation is mainly controlled by the first process yielding $P_c$(10K, 2V)=11.5%. At 10K and 2.5V, the spin relaxation is minimized yielding the highest $P_c$(10K, 2.5V)=18.8%. At 60K and 2V, the two mechanisms are coming into play thus resulting in the minimum $P_c$(60K, 2V)=5.9%. At 60K and 2.5V, the spin relaxation in the QD dominates thus giving a $P_c$(60K, 2.5V)=9.3%. It is worth mentioning that the product of $P_c$(10K, 2.5V) and $P_c$(60K, 2V) is almost equal to the product of $P_c$(10K, 2V) and $P_c$(60K, 2.5V), which provides a strong argument of the two-step spin relaxation process following Eq (2).

Thanks to the high efficient spin injection and long spin relaxation time in *p*-doped QDs at zero magnetic field, we were able to provide evidence of a two-step spin relaxation process. In previous studies, the large magnetic field (several Teslas) used for spin-LED with *in-plane* magnetized spin-injector can lead to multiple spurious effects, which modify the electron spin dynamics. Firstly, the Zeeman effect can split energy levels in the semiconductor and thus induce EL circular polarization [26]. Secondly, the magnetic field can polarize the nuclear spin and reduce the spin relaxation related to the electron-nuclear hyperfine interaction [12]. Thirdly, the large magnetic field can also affect the DP mechanism thus increasing the spin diffusion length and the spin lifetime in the semiconductor part [18]. The long spin relaxation time in our system is linked to the intrinsically long spin relaxation of time of electron within the positively charged exciton and the dynamical polarization of nuclear spins [11,12]. In our previous work [25], we have measured a nuclear field of about 200mT generated by the nuclear spin polarization at zero magnetic field after electron spin injection. This large nuclear field can protect the electron spin polarization and effectively enhance the spin relaxation time.

*6. Comparison of QD spin LED with QW spin LED*

Finally, we would like to make a comparison between QD and QW based spin LEDs having the same CoFeB/MgO PMA spin injector. The investigated QW based spin LED is characterized by a 10 nm thick undoped $In_{0.1}Ga_{0.9}As$ QW in the active region. The distance between the injector and the active region is about 100 nm instead of 80 nm for the QD based spin LED, but this effect should be negligible since the spin diffusion length is much longer than 20 nm at 10K (~3μm [50]). This allows us to compare the spin relaxation mechanism only in the different active regions. In the QW system, we have measured a 8%-10% (depending on samples) of $P_c$ at 10K under optimized bias condition [24]. By considering the value of $P_c$ measured in this work at the optimum bias (*i.e.* 19 %), we find that it is almost twice larger as that obtained for the QW based spin-LED. This comparison validates the superior performance of QD based spin-LED compared to QW based spin-LED even by using exactly the same spin injector. This superiority is mainly due to, as said previously, longer spin relaxation times for electrons in *p*-doped quantum dots compared to quantum wells, in particular when dynamic nuclear polarization occurs in CW regime, that induces a nuclear field that protects the electron spin polarization. Finally, we should also emphasize that the large circular polarization (19%) combined with very low power consumption of 15μW (with a current density about $1.4 \times 10^{-3}$ A/cm$^2$) offers a great advantage of QD-based spin-LED for future spin-optoelectronic applications. Large EL circular polarization at room temperature should be obtained using quantum dots with stronger confinement energies in order to reduce the thermal escape of carriers. It is also worth mentioning that although Be is an excellent acceptor, but its diffusivity hinders a real δ-doping [51]. Carbon would be a better candidate for δ-doping to yield a better temperature performance, which is not amphoteric and has no noticeable diffusion [52].

## IV. CONCLUSIONS

In conclusion, we have studied the spin injection and spin relaxation mechanism in *p*-doped InGaAs QD based spin-LED under zero applied magnetic field. We have measured a large circular polarization of 19% up to 100K in remanence with optimized bias condition. We have systematically studied the bias and temperature dependence of the circular polarization and evidenced a two-step spin relaxation process. The first step occurs when electrons tunnel through the MgO barrier and travel across the GaAs depletion layer. The spin relaxation is determined by the kinetic energy of the injected electrons and the transit time though the GaAs layer, as manifested by the bias dependent $P_c$. The second step occurs in the QDs prior to electron recombination with holes. The circular polarization is determined by the average doping in the QDs together with the ratio of electron lifetime and spin relaxation time, as manifested by the temperature dependent $P_c$. In addition, we have proved the superior performance of QD based spin-LEDs compared to QW based spin-LEDs having the same spin injector thanks to the longer electron spin relaxation times in *p*-doped quantum dots. This study paves the way for the development of high performance QD based spin-LEDs at room temperature without applied magnetic field, which is essential for many spin-optoelectronic applications such as optical memory element [53] and optical telecommunication with circular polarization [5].

## Acknowledgement

We thank Michel Hehn for his help on the development of ultrathin PMA CoFeB by sputtering. This work is supported by the joint French National Research Agency (ANR)-National Natural Science Foundation of China (NSFC) SISTER Project (Grants No. ANR-11-IS10-0001 and No. NNSFC 61161130527), ANR FEOrgSpin project (Grant No. ANR-18-CE24-0017) and ANR SIZMO2D project (Grant No. ANR-19-CE24-0005). We also acknowledge the PHC CAI YUANPEI 2017 (No. 38917YJ) program, ICEEL (INTER-Carnot) BlueSpinLED and ICEEL (international) SHATIPN projects as well as the French PIA project "Lorraine Universite d'Excellence" (Grant No. ANR-15-IDEX-04-LUE). Experiments were performed using equipment from the platform TUBE-Davm funded by FEDER (EU), ANR, the Region Lorraine and Grand Nancy. X. M. acknowledges the Institut Universitaire de France.

**Figures**

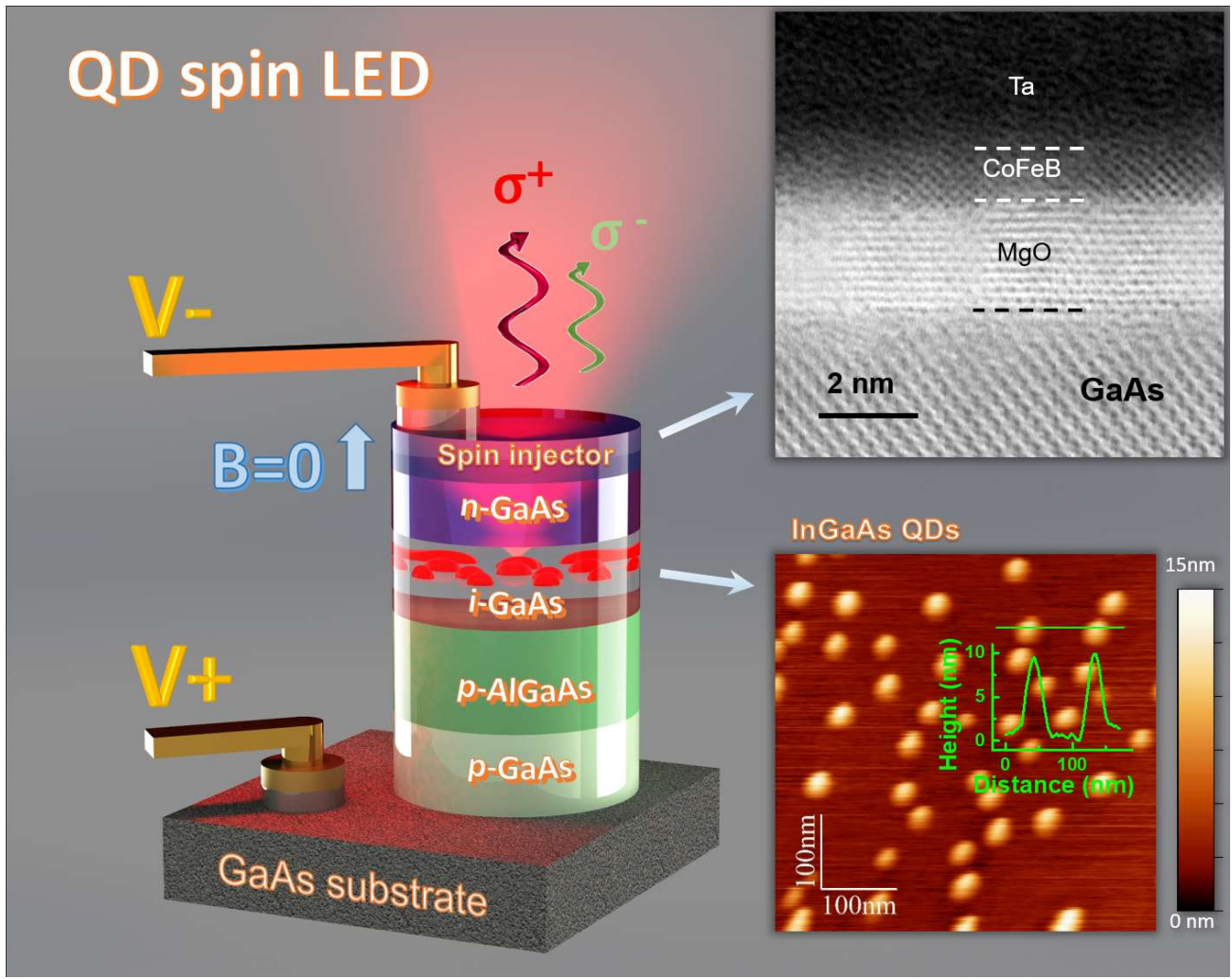

FIG. 1. Schematics of the stack structure of the QD based spin-LED which can emit circularly polarized light at zero applied magnetic field. Upper inset: HR-TEM image of the spin injector layer which contains 5 nm Ta/1.1 nm CoFeB/2.5 nm MgO. Lower inset: AFM image of the InGaAs QD layer before capping with GaAs layer. The InGaAs QDs have a density of $1.6\times10^{10}$ cm$^{-2}$. The inset of the AFM image shows the average lateral dot diameter of about 30 nm and the height of about 9 nm.

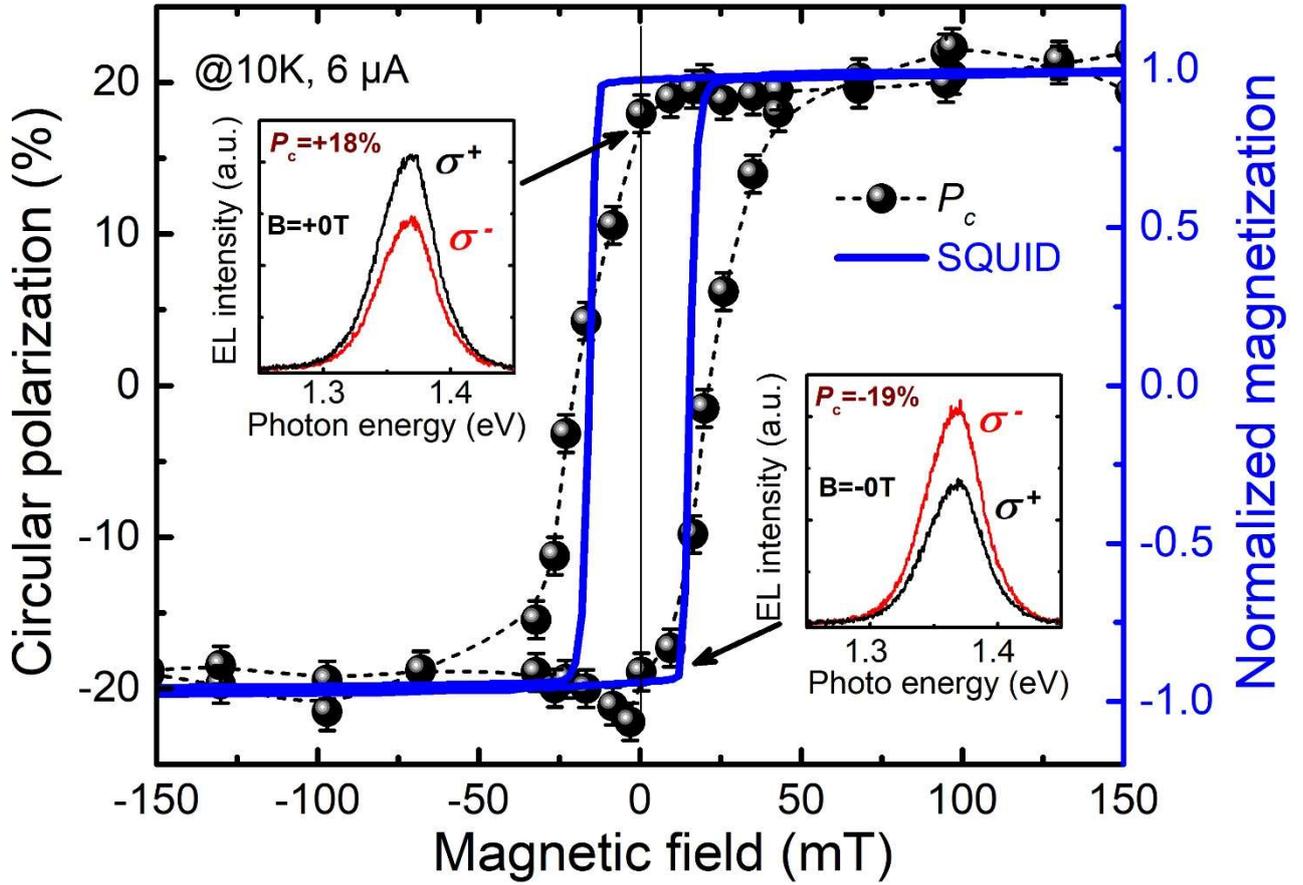

FIG. 2. Electroluminescence circular polarization $P_c$ (dashed lines with symbols) as a function of the out-of-plane magnetic field and corresponding SQUID hysteresis loop (blue solid line) measured at 10K, respectively. The insets show the electroluminescence spectra measured at zero field, where B=±0T indicate that the sample magnetization was firstly saturated by a ±0.35T field, respectively.

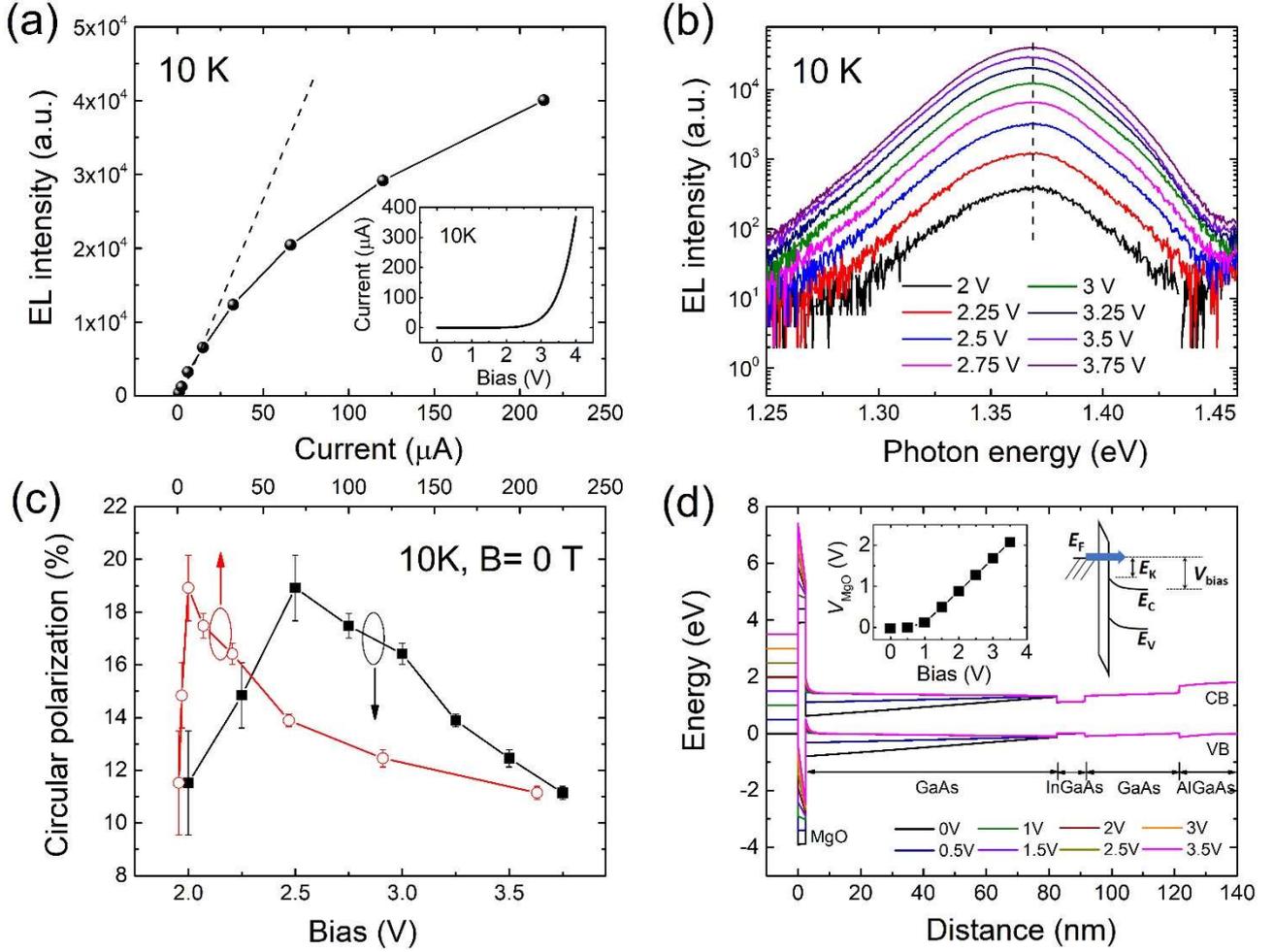

FIG. 3. (a) EL intensity as a function of the current at 10K. Inset: $I$-$V_{bias}$ curve of the sample at 10K. (b) EL spectra measured under different applied bias at 10K. The photon energy corresponding to the peak maximum (marked by dashed line) shows negligible change as a function of bias. (c) EL circular polarization as a function of the current and applied bias at 10K under zero field. (d) Band structure calculations using a one-dimensional Poisson-Schrödinger solver at 10K. Left inset: voltage drop on the MgO barrier ($V_{MgO}$) as a function of bias. Right inset: scheme of electron spin-injection from the Fermi level of the metal to the conduction band of GaAs by tunneling through the MgO barrier. Due to the bias applied on the insulating barrier, the electrons have a certain kinetic energy ($E_k$) when arriving in the GaAs layer.

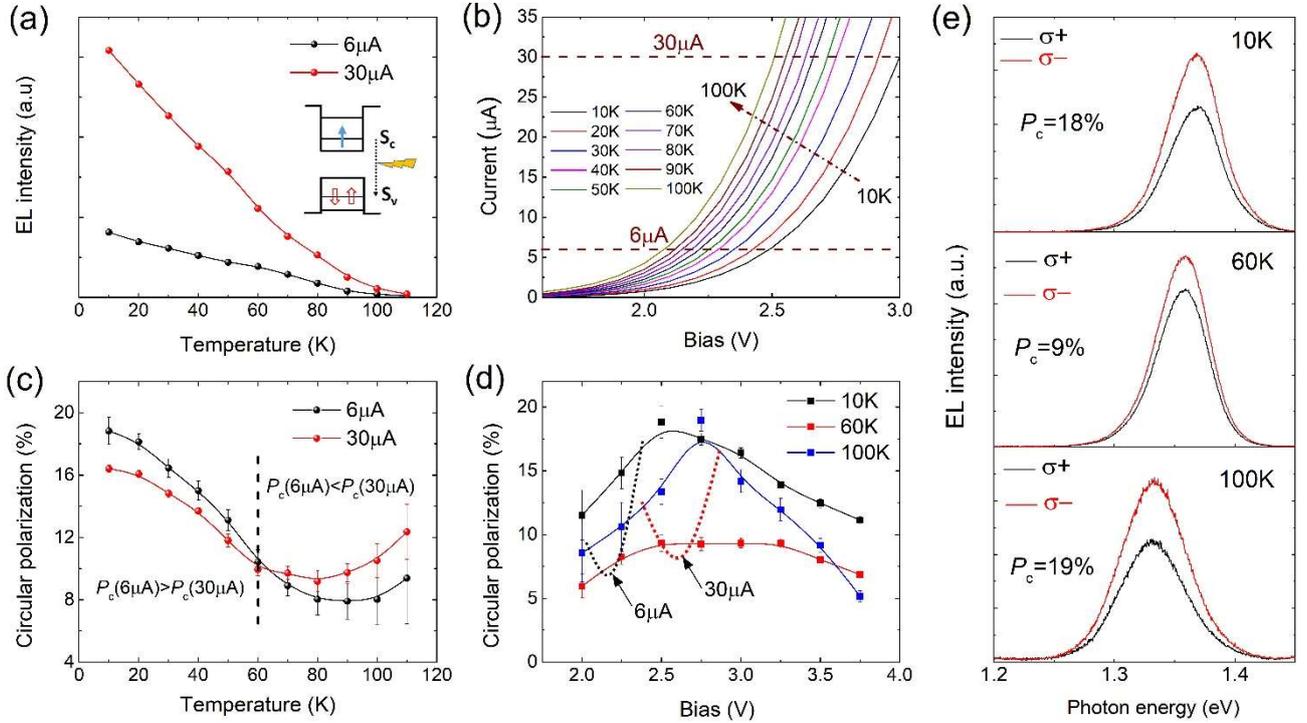

FIG. 4. (a) EL intensity as a function of temperature measured for fixed 6 µA (black dots) and 30 µA injection currents (red dots). Inset: scheme of a positively charged exciton $X^+$ formed by a spin polarized electron and two holes with opposite angular momentum projection. (b) *I-V* curves measured as a function of temperature ranging from 10 K to 100 K. (c) $P_c$ as a function of temperature with fixed injection currents of 6 µA and 30 µA at zero magnetic field. (d) $P_c$ as a function of bias at 10, 60 and 100 K at zero magnetic field. (e) Polarization resolved EL spectra at 10, 60 and 100 K at zero magnetic field with a fixed bias of 2.75 V.